\documentclass[11pt,a4paper,portrait,times]{article}
\usepackage[english]{babel}
\usepackage{url,hyperref,lineno,microtype,subcaption}
\usepackage{longtable}
\usepackage[pdftex]{graphicx} 
\usepackage[numbers]{natbib}
\usepackage{float}
\usepackage[retainorgcmds]{IEEEtrantools}
\usepackage{soul}
\usepackage{amsmath}
\usepackage{eqnarray}
\usepackage{amssymb}
\usepackage{url} 
\usepackage{lscape}
\usepackage{pdflscape}
\usepackage[utf8]{inputenc}
\usepackage[english]{babel}
\usepackage{multicol}
\usepackage{xcolor}
\usepackage{wrapfig}
\usepackage[T1]{fontenc}
\usepackage{tgbonum}
\usepackage{rotating}
\usepackage{multirow}
\usepackage[letterpaper,top=2cm,bottom=2cm,left=3cm,right=3cm,marginparwidth=1.75cm]{geometry}

\usepackage{amsmath}
\usepackage{graphicx}

\title{Transient Upstream Mesoscale Structures: Drivers of Solar-Quiet Space Weather}
\author{Primo\v{z} Kajdi\v{c}\,$^{1,*}$, Xóchitl Blanco-Cano\,$^{1}$, Lucile Turc\,$^{2}$, Martin Archer\,$^{3}$,\\ Savvas Raptis\,$^{4}$, Terry Z. Liu\,$^{5}$,  Yann Pfau-Kempf\,$^{2}$, Adrian T. LaMoury\,$^{3}$, Yufei Hao\,$^{6}$,\\  Philippe C. Escoubet$^{7}$, Nojan Omidi$^{8}$, David G. Sibeck$^{9}$, Boyi Wang$^{10}$, Hui Zhang$^{11}$\\ and Yu Lin$^{12}$}

\begin{document}
\maketitle

\begin{enumerate}
\item[$^{1}$] Departamento de Ciencias Espaciales, Instituto de Geof\' isica, Universidad Nacional Aut\'onoma de M\'exico, Mexico City, Mexico
\item[$^{2}$] Department of Physics, University of Helsinki, Helsinki, Finland
\item[$^{3}$] Blackett Laboratory, Imperial College London, London, UK
\item[$^{4}$] The Johns Hopkins University Applied Physics Laboratory, Laurel, MD, USA
\item[$^{5}$] Department of Earth, Planetary, and Space Science, University of California, Los Angeles, USA
\item[$^{6}$] Key Laboratory of Planetary Sciences, Purple Mountain Observatory, Chinese Academy of Sciences, Nanjing, People's Republic of China
\item[$^{7}$] ESA/ESTEC, Noordwijk, The Netherlands
\item[$^{8}$] Solana Scientific Inc., Solana Beach, CA,USA
\item[$^{9}$] NASA Goddard Space Flight Center, Greenbelt, MD, USA
\item[$^{10}$] Institute of Space Science and Applied Technology, Harbin Institute of Technology (Shenzhen), Shenzhen, China
\item[$^{11}$] School of Space Science and Physics, Shandong University, Weihai, China
\item[$^{12}$]  Physics Department, Auburn University, Auburn, AL, USA
\end{enumerate}

\begin{abstract}
In recent years, it has become increasingly clear that space weather disturbances can be triggered by transient upstream mesoscale structures  (TUMS), independently of the occurrence of large-scale solar wind (SW) structures, such as interplanetary coronal mass ejections and stream interaction regions. Different types of magnetospheric pulsations, transient perturbations of the geomagnetic field and auroral structures are often observed during times when SW monitors indicate quiet conditions, and have been found to be associated to TUMS. In this mini-review we describe the space weather phenomena that have been associated with four of the largest-scale and the most energetic TUMS, namely hot flow anomalies, foreshock bubbles, travelling foreshocks and foreshock compressional boundaries. The space weather phenomena associated with TUMS tend to be more localized and less intense compared to geomagnetic storms. However, the quiet time space weather may occur more often since, especially during solar minima, quiet SW periods prevail over the perturbed times.\end{abstract}

\section{Introduction}
\label{sec:TUMS-Introduction}
For decades, space weather phenomena have been thought to be strictly related to solar activity. This is mainly due to the fact that the strongest magnetospheric and ionospheric disturbances, geomagnetic storms and substorms \citep[e.g., ][]{akasofu:2021}, occur during the passage of large-scale structures (of the order of $\gtrsim$1~a.u.) in the solar wind (SW), such as interplanetary coronal mass ejections, stream interaction regions and interplanetary shocks \citep[e.g.,][]{kilpua:2017}. During such events, most extreme conditions conducive for space weather, such as large southward IMF, high speed solar wind, and large dynamic pressure (P$_{dyn}$) fluctuations, may be met. This in turn drives strong magnetopause motion and reconnectiom. 

Geomagnetic storms and substorms have been a subject of extensive research for a long time because they can interfere with our technologies by disrupting the proper functioning of, for example, electric grids, GPS signals, and artificial satellites \citep[e.g.,][]{eastwood:2017}. 

However, in recent years it has become clear that some space weather phenomena, such as bursts of large-amplitude magnetospheric ultra-low-frequency (ULF) pulsations, transient (nonperiodic) geomagnetic disturbances, auroras, etc. may occur in the absence of known space weather drivers \citep{zhang:2021}. Since their origin is not related to solar disturbances, we here refer to them as solar-quiet space weather. 

Such phenomena may be caused by transient upstream mesoscale structures (TUMS). These form in the region upstream of the bow-shock of Earth. The term mesoscale refers to their typical scale sizes ranging form $\sim$2000\,km to more than 10 Earth radii (1\,$R_\mathrm{E}\sim6400\,$km) \citep{zhang:2021}. The sizes of the largest TUMS are thus comparable to but smaller than the transverse diameter of the dayside magnetosphere \citep[$\sim$30\,R$_E$][]{tsyganenko:2014}. %

TUMS owe their existence to the collisionless bow-shock that stands in front of our planet. The bow-shock dissipates some of the SW kinetic energy by deflecting and energizing a small portion of the incident particles (electrons, ions). At its Qpar section, where the angle between the upstream IMF and the local shock normal is less than 45$^\circ$, reflected particles may escape back upstream to large distances where they coexist with the incoming SW. Such non-Maxwellian particle distributions lead to different instabilities, forming a highly perturbed foreshock region \citep{eastwood:2005a}.

The formation mechanisms for TUMS fall into three categories: (1) the interaction of IMF directional discontinuities in the SW \citep{borovsky:2008} with the bow-shock or (2) with the reflected foreshock ions and (3) due to internal foreshock processes.

The main reason why TUMS have such an impact on the near-Earth environment is the variation in magnetic field orientation and strength and the SW P$_{dyn}$ inside them which lead to modifications of the total (dynamic, thermal and magnetic) pressure impinging upon the magnetopause \citep[e.g., ][]{archer:2014}. As has been shown in the past, upstream negative and positive pressure pulses excite toroidal and poloidal mode waves in the Pc5 frequency range \citep{zong:2009, zhangXY:2010}. Even modest positive pressure pulses may also lead to an increase in temperature anisotropy of energetic protons which in turn results in ion-cyclotron instability and consequently in Pc1 magnetospheric waves \citep{olson:1983, anderson:1993b}.

P$_{dyn}$ variations have also been found to generate field aligned currents \citep[FACs,][]{araki:1994, nishimura:2016} and intensify whistler mode waves \citep[][]{li:2011b, shi:2014}. FACs can lead to electron precipitation and discrete auroras, while the intensified whistler mode waves can scatter electrons into loss cones and induce diffuse auroras.

Finally, it should be mentioned that various types of TUMS have been observed at other planets, \citep[e.g.,][]{oieroset:2001, masters:2008, slavin:2009, uritsky:2014, collinson:2012, collinson:2014, collinson:2015, collinson:2020, valek:2017, shuvalov:2019,omidi:2020b, madanian:2023b}, although their impact on the corresponding downstream regions has not been studied due to the lack of multi-spacecraft observations.

It is the purpose of this mini review to summarize the impact of the largest-scale TUMS on the near-Earth environment. In the following sections we describe such effects caused by hot flow anomalies (HFA, section~\ref{sec:TUMS-HFA}), foreshock bubbles (FB, \ref{sec:TUMS-FB}), foreshock compressional boundaries (FCB, \ref{sec:TUMS-FCB}), and travelling foreshocks (TF, \ref{sec:TUMS-ForeshockCavities}). The HFAs and TFs fall into the first category in terms of their formation mechanisms, FBs fall into the second category, while the FCBs occur due to internal foreshock processes. In section \ref{sec:TUMS-DiscussionAndConclusions} we summarize these effects while in section \ref{sec:TUMS-Future} we list some of the future tasks needed to be done in order to deepen our knowledge about the subject.

\section{Hot flow anomalies}
\label{sec:TUMS-HFA}
\begin{landscape}
    \begin{figure}[ht!]
    \centering
    \includegraphics[width=1.1\textheight]{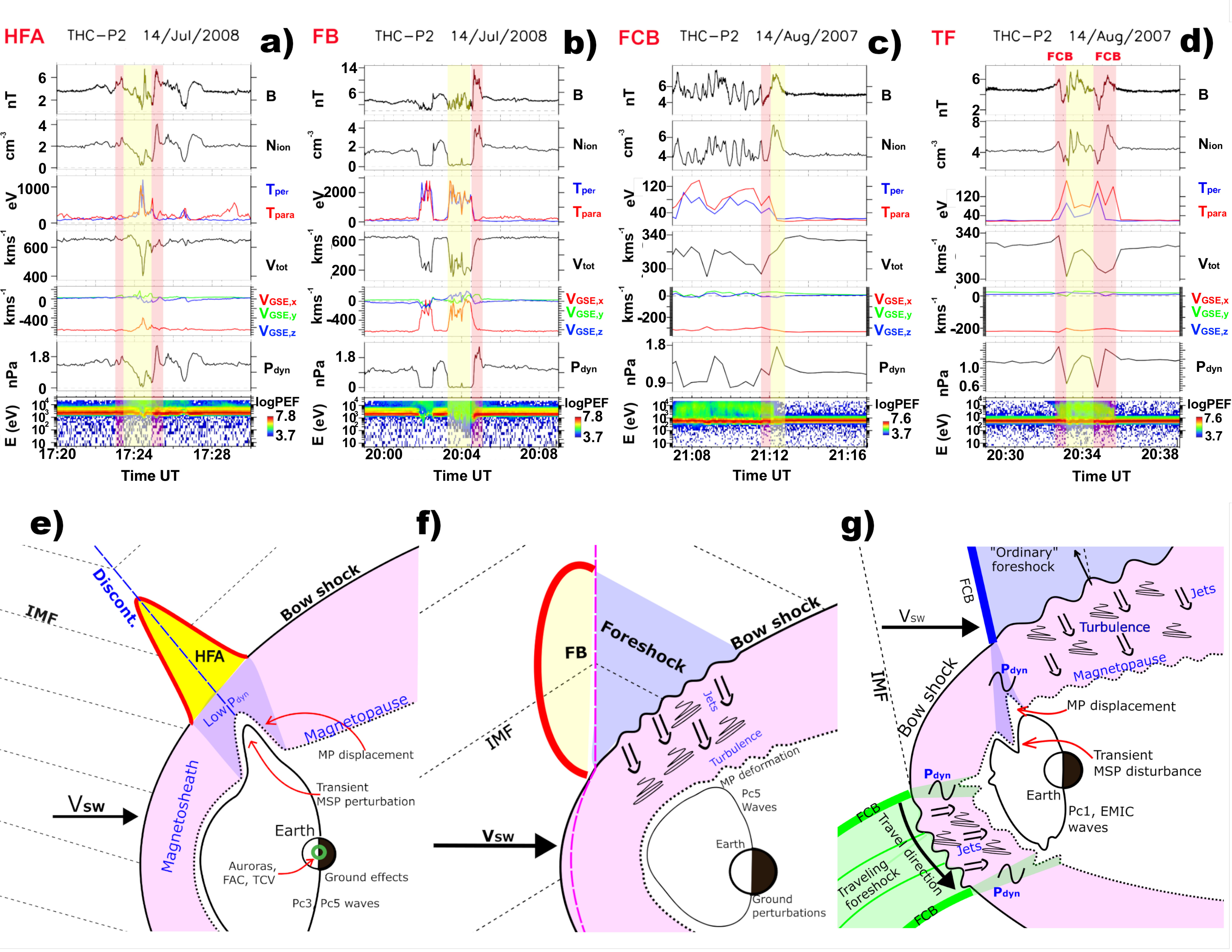}
     \caption{(a) Examples of observed (a) HFA, (b) FB, (c) FCB and (d) TF. Sketches of (e) HFA, (f) FB and (g) TF and FCB and the corresponding downstream effects. The panels a) to d) exhibit (form top to bottom) magnetic field magnitude, plasma density, parallel (blue) and perpendicular (red) ion temperatures, SW speed, SW velocity components, SW P$_{dyn}$ and ions spectra. In the case of the HFA and FB, the red shaded intervals mark rims of enhanced B and plasma density, while yellow shaded intervals mark hot cores. In the case of the FCB, the intervals shaded in red and yellow mark the B and density dip and peak, respectively. In the case of the traveling foreshock, the yellow color marks its core, while the red color marks the surrounding FCBs.}
    \label{figure1}
    \end{figure}
\end{landscape}

HFAs \citep{schwartz:1985, thomsen:1986}, form when an IMF directional discontinuity intersects the bow shock and the convection electric field ($-\mathbf{V}\times\mathbf{B}$) points towards the discontinuity's current sheet on at least one side. Their typical sizes range between 1 and 3\,$R_\mathrm{E}$ in the direction perpendicular to their current sheet, but they have been observed by \cite{chu:2017} to extend up to 7\,$R_\mathrm{E}$ upstream of the bow shock. HFAs are characterized by (see also Figure~\ref{figure1}a) central cores that contain hot plasma with flow velocities much lower than the ambient SW. The plasma flow inside HFAs is by definition highly deflected from the Sun-Earth line. The plasma density and magnetic field values in the core are lower than in the SW. The core is surrounded by a rim in which magnetic field strength and plasma density are enhanced compared to ambient SW values. An example of an HFA is shown in Figure~\ref{figure1}a.

The first geoeffective HFA was reported by \cite{sibeck:1998, sibeck:1999, borodkova:1998, sitar:1998}. An order of magnitude decrease of the P$_{dyn}$ inside the event caused the magnetopause to move outward and then inward in excess of 5\,$R_\mathrm{E}$ past Interball-1 twice within 7 minutes. Minor disturbances in geomagnetic field magnitude were observed at geosynchronous orbit by GOES-8, while Polar Ultraviolet Imager (UVI) observed a sudden brightening of the afternoon aurora, followed by a more intense transient brightening of the morning aurora.

\cite{jacobsen:2009} reported observations of extreme motion of the dawn flank magnetopause caused by an HFA. The magnetopause moved outward by at least 4.8\,$R_\mathrm{E}$ in 59\,s, implying flow speeds of up to 800\,km\,s$^{-1}$ in the direction normal to the nominal magnetopause. The transient deformation of the magnetopause generated field-aligned currents (FACs) and created travelling convection vortices \citep[e.g., ][]{glassmeier:2001} which were detected by ground magnetometers. 

Magnetopause deformation due to HFAs was also observed by \cite{safrankova:2012}. The authors reported a highly asymmetric deformation of the magnetosphere and suggested that it occurred either due to one elongated HFA or a pair of HFAs that simultaneously appeared at both flanks. On the dusk side, the deformation was very weak. On the dawn side, the magnetopause was first displaced outward from its nominal position by $\sim5\,R_\mathrm{E}$ and then inward by $\sim4\,R_\mathrm{E}$.

\cite{hartinger:2013} and \cite{shen:2018} observed HFAs that excited global Pc5  perturbations \citep[periods 150--600\,s, e.g.,][]{jacobs:1964} at the geosynchronous orbit. \cite{hartinger:2013} also reported observations of magnetopause surface modes caused by an HFA. \cite{shen:2018} demonstrated that HFAs can also generate localized magnetospheric oscillations in the Pc5 range with clear dawn-dusk asymmetry.

Several works also related passing HFAs to geomagnetic pulsations in the Pc3 range (22--100\,mHz). \cite{eastwood:2011} reported observations of an HFA associated with a type of Pc3 fluctuations whose frequency did not depend on the IMF strength, contrary to the case of Pc3 waves typically observed inside the magnetosphere \citep[e.g.][]{takahashi:1984}. Similarly, \cite{zhao:2017} reported observations of an HFA causing nearly monochromatic Pc3 ULF waves that were observed in orbit and on the ground and that exhibited characteristics of standing Alfv\'en waves. They occurred in all sectors (dawn, noon, dusk and nightside) indicating that the HFA cause a global response of the magnetosphere.

HFAs have also been shown to impact the nightside magnetosphere. This was first reported by \cite{facsko:2015} who observed an HFA remnant in the far magnetotail at $X\sim-310\,R_\mathrm{E}$. Similarly, impacts of unidentified TUMS, possibly HFAs, in the midtail magnetosheath have also been reported by \cite{wang:2018c} and \cite{liu:2020b,liu:2021c}, implying that HFAs may exhibit lifetimes of several tens of minutes.

Figure~\ref{figure1}e) summarizes the reported downstream effects of HFAs.

\section{Foreshock bubbles}
\label{sec:TUMS-FB}
Foreshock bubbles (FBs) form due to the interaction of IMF directional discontinuities with the backstreaming foreshock ions. When the they cross a discontinuity and project their velocity in the new perpendicular direction more than in the new parallel direction, the foreshock ions become more concentrated and thermalized on the upstream side of the discontinuity. Foreshock ions can easily cross rotational discontinuities (RD), since there exists a normal magnetic field component, so the ions can simply propagate along the field lines through them. At tangential discontinuities (TD), the normal magnetic field component is zero, so only ions with gyroradii larger than the TD thickness are able to cross the TDs. \citep[][]{omidi:2010, liu:2015, liu:2016, wang:2020c, wang:2021c}. Thus, stronger energy fluxes of foreshock ions are expected across RDs which may cause faster expansion of RD-driven FBs compared to TD-driven FBs.

Once ions cross the discontinuities, they undergo additional heating and start to expand against the SW, forming the bubble. FBs exhibit signatures in spacecraft data that are similar to those of HFAs (see Figure~\ref{figure1}b), namely a hot, tenuous core with low IMF strength and a rim with enhanced density and B-magnitude (see Figure~\ref{figure1}b). However, whereas HFAs commonly exhibit rims on their upstream and downstream edges, the FBs only exhibit them on their upstream side. FBs may affect the magnetopause on larger scales than HFAs since their sizes transverse to the Earth-Sun line are larger \citep[5--10\,$R_\mathrm{E}$][]{archer:2015, turner:2020}.

The first to report that FBs can be geoeffective were \cite{hartinger:2013}. The authors showed that a FB caused magnetopause undulations. Inside the magnetosphere but close to the magnetopause, the event caused variations of the North-South component of the magnetic field and similar effects were observed at geosynchronous orbit. Pc5 pulsations with similar properties as those commonly associated by the HFAs, were also observed.

\cite{archer:2015} showed that FBs have a global impact on Earth's magnetosphere. Once an FB interacts with the bow shock, magnetosheath particles are accelerated towards the intersection of the FB's current sheet with the bow shock resulting in fast, sunward flows as well as outward motion of the magnetopause. Ground-based magnetometers can detect signatures of this motion simultaneously across 7\,h of magnetic local time.

Figure~\ref{figure1}f) summarizes the reported downstream effects of FBs.

\section{Foreshock compressional boundaries}
\label{sec:TUMS-FCB}
The FCBs \citep[e.g., ][ see also Figure~\ref{figure1}c]{omidi:2009} are boundary regions that separate the highly disturbed ultra-low frequency \citep[ULF,][]{greenstadt:1995} wave foreshock from either the pristine SW or the foreshock region populated by field-aligned ion beams \citep{paschmann:1980} but not the ULF waves. FCBs are characterized by a strong compression of magnetic field magnitude and density that is followed by strong decreases of these two quantities on the foreshock side (Figure~\ref{figure1}c). These events differ somewhat from the rest of the TUMS in the sense that they are not truly transient phenomena. Models indicate that  they exist even during steady solar wind conditions and it is their motion, due to changing solar wind conditions, that has a transient impact on the magnetosphere.

\cite{hartinger:2013} described two FCBs that were observed to have an impact on the magnetopause and inside the magnetosphere. Both caused the Themis-D probe, originally located near the magnetopause on the magnetospheric side, to briefly enter the magnetosheath. Transient magnetic field and plasma density perturbations were detected throughout the dayside sector by several spacecraft located at distances corresponding to geosynchronous orbit and beyond. The timing of the perturbations observed by different spacecraft was found to be consistent with the motion of the FCB across the bow shock, in a dusk to dawn sense. Figure~\ref{figure1}g) summarizes the reported downstream effects of FCBs.

\section{Travelling foreshocks}
\label{sec:TUMS-ForeshockCavities}
TFs or foreshock cavities \citep[e.g., ][See also Figure~\ref{figure1}d]{sibeck:2002, kajdic:2017b} appear upstream of the bow shock, either in pristine SW or in the region of the ion foreshock that is not perturbed by the ULF waves. This happens when a bundle of magnetic field lines from a relatively thin magnetic flux tube, with orientation different from the background IMF, connects to the nominally quasi-perpendicular bow shock in such a way that the geometry of the section of the bow shock intersected by the flux tube is changed from quasi-perpendicular to quasi-parallel. As the flux tube is convected by the SW, its intersection with the bow shock propagates along the bow shock surface. Upstream of it, a foreshock is formed that follows this intersection. There are several ways that TFs may cause disturbances in the magnetosphere and the atmosphere (see also Figure~\ref{figure1}g).

For example, it has been reported by \cite{suvorova:2019} that two TFs drove magnetospheric ULF waves in the Pc1 frequency band. Specifically, TFs caused ground Pc1 pearl pulsations, which are amplitude-modulated Pc1 waves with a repetition period of several tens of seconds \citep[e.g.][]{jun:2014}. These pearl pulsations were observed for a long interval ($\sim$1\,hr) in the morning sector (4--8 local time, LT) and were detected at eight ground stations located at L = 3.5--7.4 (L is the distance expressed in R$_E$ at which the B-field lines cross the Earth's magnetic equator).

The same authors reported GOES-12 and THEMIS E measurements showing the Pc1 pulsations detected by the ground stations accompanied by EMIC waves in the frequency range 0.2--0.35\,Hz in the prenoon sector (7.5--12 LT) at geocentric distances between $5.8\,R_\mathrm{E}$ and $9\,R_\mathrm{E}$. The events also caused precipitation of ions with energies 30--80\,keV. Additionally, GOES-10 and 12 and THEMIS-B, -E and -D observed a transient compression of the dayside magnetosphere during which the magnetic field strength changed by up to 10\,nT and whose observed durations were of up to 5\,minutes.

Finally, \cite{sibeck:2021} and \cite{kajdic:2021b} showed that TFs are directly transmitted into the magnetosheath where they can cause the formation of enhanced P$_{dyn}$ structures, known as magnetosheath jets \citep{plaschke:2018}, in the quasi-perpendicular magnetosheath. This is the region of the magnetosheath in which the jets are rarely observed and their origins are different from those detected in the quasi-paralell magnetosheath. 

\section{Summary and discussion}
\label{sec:TUMS-DiscussionAndConclusions}
In this mini-review we discussed the reported downstream effects of the four largest-scale TUMS on the near-Earth environment. These structures may strongly affect the bow shock--magnetosheath--ionosphere system and create a wide range of space weather phenomena. It is almost certain that in the future the list of impacts of each type of TUMS will keep increasing. Table~\ref{tab:table1} summarizes explicitly reported space weather effects. 
\setlength{\tabcolsep}{6pt} 
\renewcommand{\arraystretch}{1.1} 
\begin{longtable}{m{0.6\textwidth}>{\centering}m{0.07\textwidth}>{\centering}m{0.07\textwidth}>{\centering}m{0.07\textwidth}>{\centering \arraybackslash}m{0.07\textwidth}}
\hline
& HFA & FB & FCB & TF\\
\hline
Magnetopause displacement & $\times$& $\times$& $\times$& \\
Transient geomagnetic disturbances & $\times$& $\times$& $\times$& \\
Transient magnetospheric plasma compression & & & $\times$& \\
Transient deceleration of magnetospheric plasma & & & $\times$& \\
Pc1 pulsations & & & & $\times$\\
Pc3 pulsations & $\times$& & & \\
Pc5 pulsations & $\times$& $\times$& & \\
Magnetospheric EMIC waves & & & & $\times$\\
Ion precipitation & & & & $\times$\\ 
Field-aligned currents & $\times$& & & \\
Travelling convection vortices & $\times$& & & \\
Ground magnetic field perturbations & $\times$& $\times$& & $\times$\\
Auroral brightenings & $\times$& & & \\
Magnetosheath jets &   &   & & $\times$\\
\hline
\caption{Transient upstream mesoscale structures and observed downstream effects.}
\label{tab:table1}
\end{longtable}

We still do not understand all the mechanisms by which different TUMS affect the regions downstream of the bow shock.

For example, we do not know how the monochromatic Pc3 fluctuations are caused by HFAs. One possibility is that shocks that sometimes form at the HFAs and FBs steepened edges, drive their own foreshocks with ULF fluctuations which eventually perturb the magnetosphere, similar to the ULF waves in the terrestrial foreshock \citep[e.g.,][]{engebretson:1987, turc:2023}. Turbulence and waves in the cores of these structures \citep{zhang:2010, kovacs:2014} could also be the cause.

Another possible effect that has not yet been well studied is that TUMS associated enhancements of P$_{dyn}$ could lead to impulsive penetration of mass into the magnetosphere \citep{dmitriev:2015}. Modification of the IMF upstream and in the magnetosheath could also result in magnetopause reconnection \citep{hietala:2018}.

These effects could be caused by TUMS associated magnetosheath jets \citep{plaschke:2018}. It has been shown by \cite{sibeck:2021} and \cite{kajdic:2021b} that the TFs transmitted into the magnetosheath can be a source of these jets downstream of the quasi-perpendicular bow-shock. \cite{nykyri:2019,dmitriev:2023}  have demonstrated that magnetosheath jets can be geoeffective and can act as a vector for coupling TUMS and foreshock processes to the magnetopause and ionosphere.
 
 To make matter worse, certain types of TUMS can contain another type of upstream mesoscale structures. The latter is most evident in the case of TFs that often contain FCBs at their edges \citep[][]{kajdic:2017b}. Moreover, TFs exhibit other phenomena that are also observed inside the ``regular'' foreshock, such as ULF waves, shocklets, foreshock cavitons, etc.

\section{Future work}
\label{sec:TUMS-Future}
It is clear that our knowledge of how exactly TUMS interact with the bow shock and the regions downstream of it is still limited.  Future investigations should include more multi-point observations of individual events with spacecraft in different regions (upstream of the bow shock, magnetosheath, magnetosphere, ground observations). These should be accompanied by local and global numerical simulations. There are numerous tasks in the ``to do'' list:
\begin{itemize}
    \item Study of the microphysics in the cores and the boundary regions of the TUMS, i.e. possible generation of ULF waves and turbulence, magnetic reconnection, particle heating and acceleration.
    \item Study of the impact of foreshock cavitons and spontaneous hot flow anomalies on the regions downstream of the terrestrial bow-shock.
    \item Comparison study of properties and impact if FBs formed by rotational versus tangential discontinuities.
    \item Detailed investigations of the impact of the TUMS on the bow shock. Do TUMS cause shock erosion, its additional rippling and what are the downstream consequences of these processes?
    \item Studies of the TUMS's substructure and the physical processes leading to it.
    \item Direct observational confirmation between the TFs and the magnetosheath jets and Pc3--4 waves in the magnetosphere.
    \item Statistical study that would reveal the relative importance of travelling versus the ``regular'' foreshocks for the production of magnetosheath jets and Pc3--4 waves.
    \item Determine the impact of each type of TUMS on the nightside magnetosphere. For example, can they trigger substorms?
    \item Test whether energetic particles accelerated in the foreshock and TUMS can enter into the magnetosphere (across the magnetopause or through the cusp) and become geoeffective.
    \item Quantify the energy input from TUMS into the magnetosphere in comparison with typical solar wind drivers.
    \item Determine the role of TUMS during storm time (e.g., enhance magnetospheric ULF waves and thus modulate radiation belt particles).
    \item Determine how HFAs excite the Pc3 waves and whether they can also be caused by FBs.
    \item Determine the impact of TUMS on the near-planetary environment at other planets. One such opportunity will emerge with the dual orbiter BepiColombo mission at Mercury.
\end{itemize}

Such tasks require multi-point spacecraft observations as well as 3D physically scaled global numeric models that go beyond the fluid description of plasma. Currently, numerous in-situ and ground based observations are available as well as the required kinetic simulation assets that will make addressing these tasks possible.

\section{Acknowledgments}
This work was supported by the International Space Science Institute (ISSI) through ISSI International Team project \#555. The ClWeb (https://clweb.irap.omp.eu/) tool was used for visualizing the data and producing some of the figures. PK's work was supported by the DGAPA PAPIIT through the IN100424 grant. XBC acknowledges DGAPA PAPIIT grant IN106724. The work of LT was supported by the Research Council of Finland (grant number 322544). SR acknowledges funding from NASA DRIVE Science Center for Geospace Storms (CGS) — 80NSSC22M0163. DGS's work was supported by NASA’s LWS TR\&T program. MOA was supported by UKRI (STFC/EPSRC) Stephen Hawking Fellowship EP/T01735X/1 and UKRI Future Leaders Fellowship MR/X034704/1. YP acknowledges Academy of Finland grant no. 339756. ATL was supported by Royal Society awards URF\textbackslash R1\textbackslash180671 and RGF\textbackslash EA\textbackslash 181090. HZ acknowledges National Natural Science Foundation of China grant 42330202.

\newpage

\bibliographystyle{alpha}

\end{document}